\documentclass{emulateapj}

\newcommand{\sigs}{\sigma_s}

\newcommand{\meanrho}{\rho_0}

\newcommand{\means}{s_0}

\newcommand{\mach}{\mathcal{M}}
\newcommand{\macha}{\mathcal{M}_\mathrm{A}}

\newcommand{\g}{\mathrm{g}}

\newcommand{\s}{\mathrm{s}}

\newcommand{\cs}{c_{\rm s}}

\usepackage[dvipsnames]{xcolor}

\usepackage{verbatim}
\usepackage{amsmath}
\usepackage{multirow}
\usepackage{relsize}
\usepackage{graphicx}

\setcitestyle{aysep={}} 
\shorttitle{Protostellar Feedback and the Density PDF}
\shortauthors{Appel et al.}

\begin{document}

\title{The Effects of Magnetic Fields and Outflow Feedback on the Shape and Evolution of the Density PDF in Turbulent Star-Forming Clouds}

\author{Sabrina M. Appel}
\email{appel@physics.rutgers.edu}
\affiliation{Department of Physics and Astronomy, 
Rutgers University,
136 Frelinghuysen Rd., 
Piscataway, NJ 08854, USA}

\author{Blakesley Burkhart}
\affiliation{Department of Physics and Astronomy, 
Rutgers University,
136 Frelinghuysen Rd., 
Piscataway, NJ 08854, USA}
\affiliation{Center for Computational Astrophysics, 
Flatiron Institute, 
162 Fifth Avenue, 
New York, NY 10010, USA}

\author{Vadim A. Semenov}
\affiliation{Center for Astrophysics $|$ Harvard \& Smithsonian, 
60 Garden St., 
Cambridge, MA 02138, USA}

\author{Christoph Federrath}
\affiliation{Research School of Astronomy and Astrophysics, 
The Australian National University, 
Canberra, ACT 2611, 
Australia}

\author{Anna L. Rosen}
\affiliation{Center for Astrophysics $|$ Harvard \& Smithsonian, 
60 Garden St., 
Cambridge, MA 02138, USA}

\begin{abstract}

Using a suite of 3D hydrodynamical simulations of star-forming molecular clouds, we investigate how the density probability distribution function (PDF) changes when including gravity, turbulence, magnetic fields, and protostellar outflows and heating. 
We find that the density PDF is not lognormal when outflows and self-gravity are considered.
Self-gravity produces a power-law tail at high densities and the inclusion of stellar feedback from protostellar outflows and heating produces significant time-varying deviations from a lognormal distribution at the low densities.
The simulation with outflows has an excess of diffuse gas compared to the simulations without outflows, exhibits increased average sonic Mach number, and maintains a slower star formation rate over the entire duration of the run. 
We study the mass transfer between the diffuse gas in the lognormal peak of the PDF, the collapsing gas in the power-law tail, and the stars. 
We find that the mass fraction in the power-law tail is constant, such that the stars form out of the power-law gas at the same rate at which the gas from the lognormal part replenishes the power-law. 
We find that turbulence does not provide significant support in the dense gas associated with the power-law tail.  
When including outflows and magnetic fields in addition to driven turbulence, the rate of mass transfer from the lognormal to the power-law, and then to the stars, becomes significantly slower, resulting in slower star formation rates and longer depletion times.

\end{abstract}


\section{Introduction} 
\label{sec:intro}

Star formation takes place in dense and cold giant molecular clouds (GMCs) that are subject to magnetized supersonic turbulent motions \citep[e.g.,][]{Padoan1997,Mckee_Ostriker2007,Laz09rev,KE2012,Myers14a,KrumholzBurkhart2018}.   
Star formation is, in part, controlled by non-linear fluid dynamics, and models of star formation focus on understanding the roles that self-gravity, magnetohydrodynamic (MHD) turbulence, and galactic environment play in the formation of dense collapsing gas that may form stars \citep[e.g.,][]{Padoan12a,Collins12a,Burkhart2017}. 
For individual star-forming clouds, a common approach to modeling the relevant physics is to investigate the distribution of gas via a density probability distribution function (PDF) analysis \citep[e.g.,][see \citealt{Padoan14a} for a review]{KrumholzMcKee2005,PadoanNordlund2011,FederrathKlessen2012,Burkhart2018}. 
Models for star formation that use a density PDF have been used to explain a range of phenomena including the mass distribution of cores and the stellar initial mass function \citep[e.g.,][]{HennebelleChabrier2008,HennebelleChabrier2009,HennebellChabrier2011,Hopkins2012}. 
Such models can also be used as subgrid models in galaxy formation simulations to model the dependence of local star formation on the dynamical state of the gas \citep{Braun2015,Semenov2016,Semenov2021,Li2017,Trebitsch2017,Lupi2018,Gensior2020,Kretschmer2020,Kretschmer2021,Olsen2021}.

The shape of the density PDF in GMCs is expected to be lognormal when supersonic turbulence dominates the gas dynamics and to transition to a power-law as gravitational contraction takes over at high densities \citep[see e.g.,][]{Klessen2000,Kritsuk+2011,Collins12a,Federrath13a,JaupartChabrier2020,Khullar2021}. 
This shape has been seen in both observations (i.e., in column density) and in numerical simulations \citep[both in density and projected column density; see e.g.,][]{VazquezSemadeni2001, Wada07a,Ossenkopf-Okada2016, Veltchev2019}.
The lognormal form of the column density PDF describes the behavior of diffuse HI and ionized gas as well as some molecular clouds that are not forming massive stars \citep{Hill2008,burkhart10, Kainulainen13a,Schneider+2015}. 
The dense gas in molecular clouds (i.e., extinctions greater than $A_v >1$, corresponding to $n>10^3\;$cm$^{-3}$) is predominantly found to have a power-law PDF rather than a lognormal PDF \citep{Collins12a,Girichidis2014,MyersP2015,schneider15,Burkhart2017,Mocz2017,MyersP2017,Alves2017AA,Chen2017,Kainulainen2017,Chen2018}. 

Most analytic models of star formation have used only a lognormal form for the density PDF, where the width of the distribution is set by the properties of MHD turbulence, including the sonic Mach number and the type of driving \citep[e.g.,][]{KrumholzMcKee2005,FederrathKlessen2012,Salim15a}. 
The star formation rates are obtained by integrating the PDF past a predefined critical density for collapse\footnote{The critical density is determined by the competition of supportive terms (e.g., large-scale turbulent motions, thermal pressure, and magnetic fields) with compressive terms such as gravity and shocks. This density varies in the literature \citep[e.g.,][]{KrumholzMcKee2005,PadoanNordlund2011,HennebellChabrier2011,FederrathKlessen2012}.}. 
However, models that use a lognormal density PDF lack a time-varying treatment of stellar feedback (e.g., stellar winds, ionizing radiation, radiation pressure, and collimated outflows or jets) and neglect the evolution of the PDF shape under the influence of self-gravity. 

To remedy this, such models modify the star formation rate (SFR) by an efficiency factor, such as with a constant parameter $\epsilon_0$ or $\epsilon_{\rm core}$ that is due to stellar feedback, since feedback processes, along with magnetic fields, are now understood to be key ingredients for the low efficiency of star formation \citep [see e.g.,][]{Rosen2014,Federrath2015,Grudic+2018,RosenKrumholz2020,Rosen+2020}. 
As such, analytical models of star formation that assume a lognormal form for the density PDF are only able to take into account the important aspects of feedback in simple ad-hoc ways that reduce the overall star formation rates and efficiencies down to the observed values. 
Additionally, because the lognormal form is set only by turbulence, it neglects the effects of self-gravity on the density distribution, e.g., the power-law tail\footnote{Gravity can be accounted for in a `multi-freefall' way that accounts for the varying freefall times of gas at different densities by weighting the integral with a freefall-time factor \citep[see e.g.,][]{HennebellChabrier2011,FederrathKlessen2012}.
However, those works still use lognormal distributions at all densities.}.
The static nature of lognormal PDF models neglects the impact of gas cycling from warm diffuse gas to cold dense star-forming gas in the ISM, which may be critical for setting long depletion times, as well as the star formation efficiency \citep{Semenov2017,Semenov2018}.

Motivated by these studies, \cite{Burkhart2018} presented an analytic model for star formation in a gravo-turbulent medium based on a piecewise lognormal and power-law density PDF that can allow for mass cycling. 
The model presented in \cite{Burkhart2018} and \citet{BurkhartMocz2019} accounts for self-gravity and stellar feedback by including a power-law tail on the high-density end of the density PDF and allowing the form of the PDF to be time variable. 
This model estimates the SFE by taking the ratio of the mass of star-forming gas in the power-law tail to the total gas in the cloud.  
The model has thus far been tested on MHD gravo-turbulent simulations that lacked stellar feedback \citep{BurkhartMocz2019, Khullar2021}.

Few studies have focused on understanding how mass transfers between different parts of the density PDF and how the overall shape of the PDF is affected by the inclusion of stellar feedback. 
In this work, we analyze the density PDFs of a suite of hydrodynamical simulations of star formation within a molecular cloud that progressively adds physics starting from only self-gravity, then adding turbulence, magnetic fields, and finally including stellar feedback from protostellar jets and slower but wider disk winds\footnote{The outflow model used in this paper uses a combination of a fast collimated component, which represents jets launched from the inner part of the accretion disk, and a slower, wide-angle component, which represents disk outflows and/or entrained material. For the remainder of this paper, we will refer to the combination of the collimated component and the slower, wider component as protostellar outflows or, simply, outflows.}. 
Collimated protostellar jets are magnetically launched by the winding-up of the magnetic field in the accretion disk, following a magneto-centrifugal mechanism \citep[][]{Blandford82a,Shu+1988,Pelletier+1992,bontemps96,LyndenBell2003,Maud+2015,Kolligan+2018,Rosen+2020}. 
Jets and disk winds are well known to play a significant role in setting the structure of their host molecular cloud and in setting star formation rates \citep[see e.g.,][]{Federrath+2014,Rosen+2020}. 
In simulations, protostellar outflows are often included via subgrid models, such as the model described in \citet{Federrath+2014}, and used in this paper. 

Using these simulations, we investigate how the density PDFs of star-forming molecular clouds evolve when we separately study the effects of gravity, turbulence, magnetic fields, and protostellar outflows. 
Our goal is to understand how different physical processes influence the density PDF and the movement of material from diffuse gas into stars. 
We also analyze how different portions of the density PDF change over time and how the mass flow of the gas connects to the change in the star formation rate with time. 

This paper is organized as follows: 
In Section~\ref{sec:numerics}, we describe the setup of the simulations used in this paper, and in Section~\ref{sec:pdfs}, we review analytic models for density PDFs. 
In Section~\ref{sec:models}, we discuss our fits to the simulated density PDFs and detail how mass flows between different sections of the PDF and into stars.
In Section~\ref{sec:discussion}, we discuss the implications of our results, and in Section~\ref{sec:conclusion}, we summarize our work.

\section{Simulations and Numerical Parameters} \label{sec:numerics}

\begin{deluxetable*}{ccccccccccc}[ht]
\tabletypesize{\footnotesize}
\tablecaption{Summary of key simulation parameters
\label{tab:sims}}
\tablecolumns{5}
\label{tab:sims}
\tablewidth{0pt}
\tablehead{
\colhead{Simulation} &
\colhead{Gravity?} &
\colhead{$\sigma_{v}$ ($\mathrm{km \, s}^{-1}$)} &
\colhead{$b$} &
\colhead{$\mach_s$} &
\colhead{B ($\mu$G)} &
\colhead{$\macha$} &
\colhead{Outflows?} &
\colhead{$\rho_t $ ($\mathrm{g \,  cm}^{-3}$)} &
\colhead{$N_{\rm res}^{3}$} &
\colhead{Reference}
}
\startdata
 \textsc{Gravity}    &  yes & N/A & N/A & N/A  & N/A & N/A & no & N/A  & $1024^3$ & \cite{Federrath2016} \\
 \textsc{Turbulence}  &   yes & $1 $ & 0.4 & 5    & N/A  & N/A & no & $1.64 \times 10^{-20}$ & $1024^3$ & \cite{Federrath2015} \\
 \textsc{B-Fields} &  yes & $1$ & 0.4 & 5    & 10  & 2.0 & no & $1.64 \times 10^{-20}$ & $1024^3$ & \cite{Federrath2015} \\
 \textsc{All + Outflows}   &  yes & $1 $ & 0.4 & 5  & 10 & 2.0 & yes & $1.64 \times 10^{-20}$ & $2048^3$ &  \cite{Federrathetal2017} \\
 \enddata
\vspace{-0.3cm}
\tablecomments{A summary of the simulations and key parameters: $\sigma_{v}$ is the velocity dispersion, $b$ is the turbulent driving parameter, $\mach_s$ is the sonic Mach number, B is the strength of the magnetic fields, $\macha$ is the Alfv\'{e}nic Mach number, and $\rho_t $ is the calculated reference transition density. $N_{\rm res}^{3}$ shows the maximum effective grid resolution, and we list the reference for the paper in which each simulation was first presented. Note that the simulation with outflows also includes protostellar heating feedback. The initial mean volume density of each simulation is $\rho_0 = 3.28 \times 10^{-21}$ g cm $^{-3}$.}
\end{deluxetable*}

\begin{figure*}[ht!]
\includegraphics[width = \linewidth]{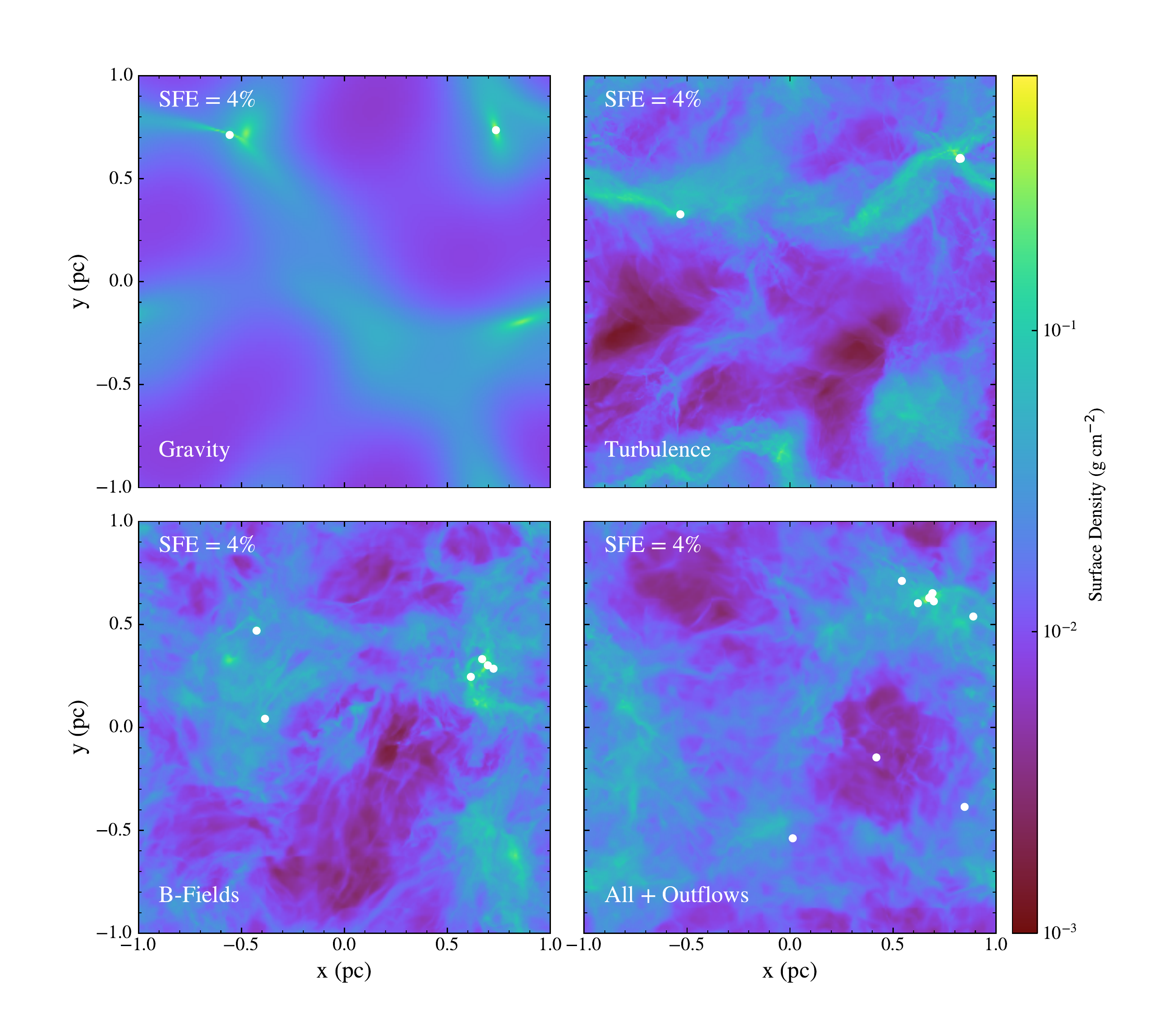}
\caption{Density projection plots along the z-axis for each of the four simulations described in Table~\ref{tab:sims}. Following the order in Table~\ref{tab:sims}, each simulation includes progressively more physical processes. In this paper, we analyze a series of snapshots for each simulation corresponding to increasing integrated star formation efficiency. The projection plots shown here are of the 4\% integrated star formation efficiency snapshots. White circles show the x-y positions of the sink particles.
\label{fig:projection}}
\end{figure*}

For our analysis, we use a suite of four simulations that model a star-forming region within a molecular cloud. 
These four simulations are drawn from \cite{Federrath2015,Federrath2016} and \cite{Federrathetal2017} and include varying physical processes, such as self-gravity, turbulence, magnetic fields, and protostellar outflows.
All four simulations were performed using the FLASH hydrodynamics code, a publicly available adaptive mesh hydrodynamics code \citep{Fryxell2000}. 
FLASH solves the fully compressible MHD equations using adaptive mesh refinement (AMR) and can include many inter-operable modules. 
The simulations presented here use a Godunov-type method with the second-order, 5-wave approximate HLL5R Riemann solver \citep[][]{Waagan+2011}. 

The first simulation (\textsc{Gravity}) includes only self-gravity with an initial Gaussian random density distribution and zero initial velocities \citep[see][for details]{Federrath2016}. 
The second simulation (\textsc{Turbulence}) initializes with uniform density and drives turbulence for two eddy-turnover times ($t_{\rm turnover} = L/(2 \sigma_{v}) \approx 0.98$~Myr, where $L = 2$~pc is the box size and $\sigma_{v} = 1$ km s$^{-1}$ is the velocity dispersion on the driving scale $L/2$), before turning on self-gravity. 
Turbulence is continually driven at half the box size throughout the run to maintain the sonic Mach number close to $\mach_s=5$ and with a natural mix of forcing modes, such that the resulting turbulence driving parameter is $b \sim 0.4$ \citep[see the turbulence driving method developed in][]{Federrath10b}. 
The third simulation (\textsc{B-Fields}) is identical to the \textsc{Turbulence} run, but adds magnetic fields with an Alfv{\'e}n Mach number of $\macha=2$.  
Both the \textsc{Turbulence} and \textsc{B-Fields} runs were first presented in \cite{Federrath2015}.
The fourth simulation (\textsc{All + Outflows}) is similar to the \textsc{B-Fields} run but adds stellar feedback in the form of protostellar outflows and protostellar heating feedback, as described in \cite{Federrathetal2017}. 
Table~\ref{tab:sims} lists the physical parameters of each simulation, and Fig.~\ref{fig:projection} shows density projections for each of these four simulations.

We ran an additional simulation without gravity (\textsc{No Gravity}, not shown in Table~\ref{tab:sims}) to demonstrate the effect of self-gravity on the density PDF. 
This simulation has the same initial conditions as the \textsc{Turbulence} and \textsc{B-Fields} simulations in Table~\ref{tab:sims} and includes both turbulence and magnetic fields, but does not include gravity and therefore does not produce sink particles.
We use one snapshot from this simulation only to compare the density PDF with the other four simulations and do not use it in the rest of the analysis.

All of the simulations have a box size of 2~pc and a total mass of $M_{\rm init}=388 \, M_{\odot}$, resulting in a mean density of $\rho_0 = 3.28 \times 10^{-21}$ g cm $^{-3}$. 
Periodic boundary conditions are used for all of the simulations.
The virial parameter (the ratio of twice the kinetic energy to the gravitational energy) is $\alpha_{\rm vir} = 1.0$.
The simulations that include magnetic fields have a field strength of $B = 10 \, \mu$G, an Alfv\'{e}n Mach number of $\macha = 2.0$, and a plasma beta parameter (representing the ratio of thermal and magnetic pressure) of $\beta = 0.33$ \citep{Federrath2015}.
All of the simulations are isothermal except for the \textsc{All + Outflows} simulation which is initially isothermal but allows for protostellar heating as implemented in \citet{Federrathetal2017}. No cooling is modeled.

To account for the formation of stars, the simulations use sink particles as described in \cite{Federrath10a} and \cite{Federrath+2014}. 
Sink particles are formed when a local region undergoes gravitational collapse and exceeds a threshold density, $\rho_{\rm sink}$. 
The \textsc{Gravity}, \textsc{Turbulence}, and \textsc{B-Fields} simulations form sink particles at $s_{\rm sink} = \ln (\rho_{\rm sink}/\rho_{0}) = 8.75$. 
The \textsc{All + Outflows} simulation, which has a higher maximum resolution than the other simulations, forms sink particles at $s_{\rm sink} = \ln (\rho_{\rm sink}/\rho_{0}) = 10.14$.
Sink particles only form in regions that are maximally refined ($\rho_{\rm sink}$ is greater than all of the grid level refinement density thresholds). 
The sink radius is set to 2.5 grid cell lengths (of the maximally refined cells) to avoid artificial fragmentation following the Jeans criterion from \citet{Truelove98a}. 
On all other AMR levels, the refinement criterion is set to refine the local Jeans length by at least 32~grid cells, except for the \textsc{All + Outflow} model, which resolves the Jeans length by at least 16~cells \citep{Federrath+2011}.
Sink particles continue to accrete gas from any cells within the accretion radius of the sink particle that have exceeded $\rho_{\rm sink}$, and are gravitationally bound and converging towards the sink \citep{Federrath10a}.

The \textsc{All + Outflow} simulation uses a custom module for implementing two-component jet feedback as described in \citet{Federrath+2014}. 
This module consists of a wide-angle low-speed component and a collimated high-speed component, motivated by observations. 
The outflow module has been physically calibrated based on theoretical models of jet launching, dedicated numerical simulations of single accretion disks with jets, and observations.
For more details about the sink particles used in these simulations and the protostellar outflow prescription, see e.g., Fig.~2 of \cite{Federrath+2014}. 
See \cite{Federrath2017} for further details about the protostellar heating feedback. 

\begin{figure}[tb!]
\includegraphics[width = \linewidth]{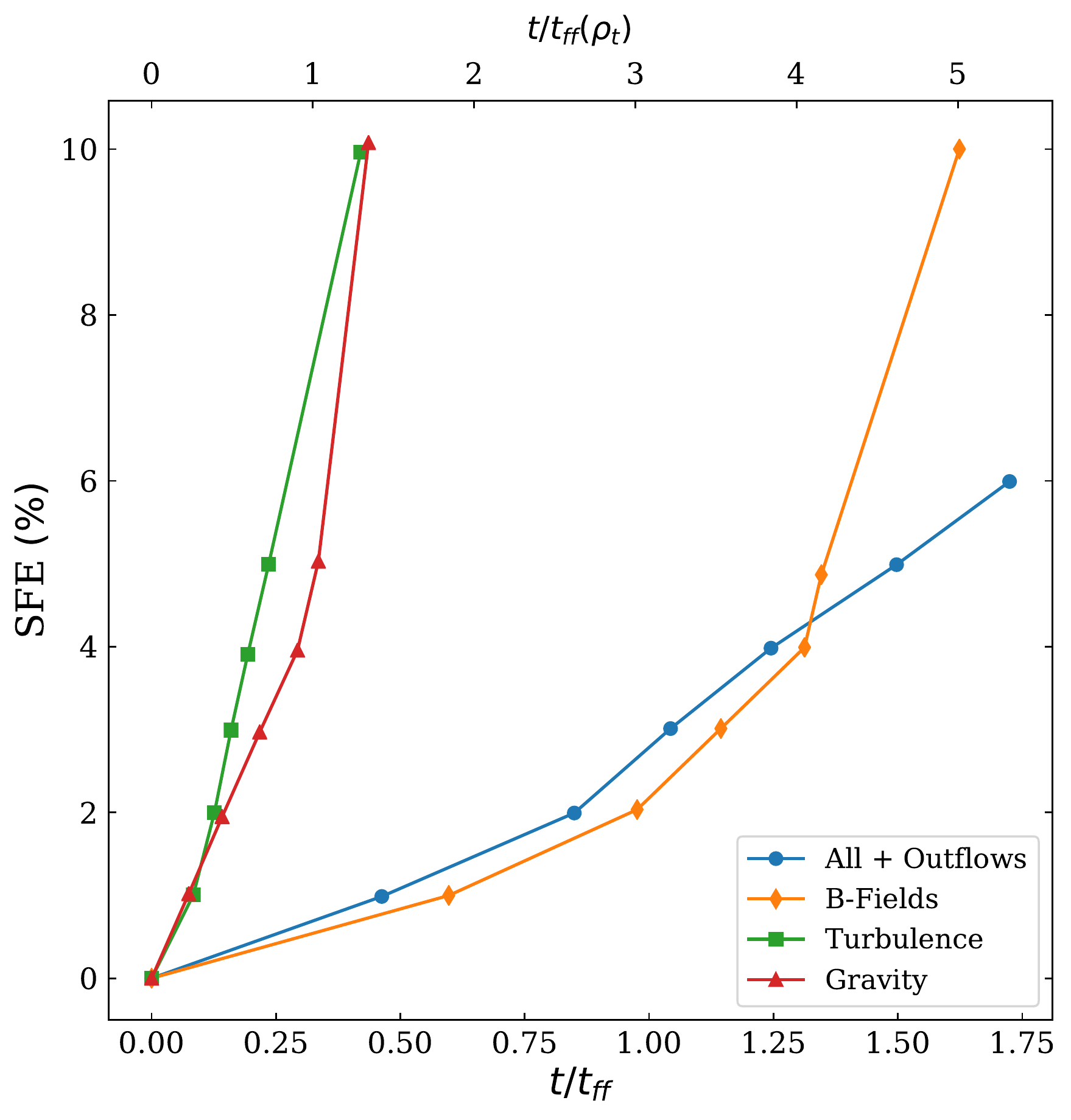}
\caption{The integrated star formation efficiency (SFE) is shown as a function of time for each simulation. The integrated SFE is calculated as $M_*/M_{\rm init}$ and represents the ratio of the stellar mass and the total initial mass of the simulation, which is equivalent to the sum of the gas mass and stellar mass for a given snapshot. SFE is shown as a function of time, where time is measured in freefall times of the average density ($t_{\text{ff}} \approx 1.16$ Myr) on the lower axis and in freefall times of the reference transition density (discussed below, see e.g., Eq.~\ref{eq:s_t}) on the upper axis ($t_{\text{ff}}(\rho_{t}) \approx 0.38$ Myr). Time is measured since the $0\%$ snapshot, i.e., right before the first sink particle is formed.  \label{fig:intSFE}}
\end{figure}

As would be expected, the four simulations evolve very differently (see, for example, the projection plots in Fig.~\ref{fig:projection}). 
As demonstrated in \citet{Federrath2015} the differing physics produces vastly different instantaneous star formation rates. 
In addition, the first sink particle is produced at different times in each simulation.  
The output files are saved based on the integrated star formation efficiency (how much of the gas has been converted into stars), thus each of the simulation snapshots corresponds to a different physical time. 
Figure~\ref{fig:intSFE} shows the integrated SFE ($M_{*}/M_{\rm init}$) as a function of time for each of the simulations. 
Note the similarity of the SFE evolution between the \textsc{Turbulence} run and \textsc{Gravity} run is somewhat misleading as these two simulations have very different initial conditions for their density fields. 
This does not imply that turbulence does not play a role in setting the SFR. 
In fact, for the same initial condition density field, \citet{Federrath2015}
shows that the addition of turbulence reduces the SFR per free fall time by a factor of about 2--3. 

The first snapshot is from just before the first sink particle is formed for each simulation -- this is the SFE~$=$~0\% snapshot and we define this to be the $t=0$ point in our analysis (see e.g., Fig.~\ref{fig:intSFE}). 
Note that the amount of time that passes between when turbulence is fully developed and self-gravity is turned on and when the first sink particle is formed is different for each simulation. 
Since the latter point is where we set $t=0$, we expect gravity to have already established a power-law tail by $t=0$. 
We choose this as our $t=0$ point since we are interested in the evolution of the cloud during star formation and are not currently considering the preceding evolution.
Subsequent snapshots may be referred to by their integrated SFE, which can be seen as a proxy for time as a higher SFE refers to a more evolved snapshot.
For example, the 4\% snapshot is when 4\% of the gas mass has been converted into stellar mass (this is the snapshot shown in Fig.~\ref{fig:projection}).

For each of the simulations, we consider the SFE~$=0$\%, 1\%, 2\%, 3\%, 4\%, 5\%, and 10\% snapshots.
Due to a much longer computational run time, the SFE~$=$~10\% snapshot for the \textsc{All + Outflows} simulation was not available and we instead consider a SFE~$=$~6\% snapshot.

\section{Analytic Description of Density PDFs} \label{sec:pdfs}

\subsection{Pure Lognormal Density PDFs}\label{sec:LNmodels}

A standard approach for modeling star formation at the cloud scale is to assume a time-independent, lognormal density PDF, which is the expected density distribution for supersonic turbulent molecular clouds \citep[e.g.,][]{Vazquez-Semadeni1994,Vazquez-Semadeni95a,Padoan1997,Scalo98a,Kravtsov03a,Robertson2008,HennebelleChabrier2008,Price10b,Collins12a,Burkhart12,Hopkins2013,Walch13a,KrumholzMcKee2005,Molina2012,FederrathKlessen2012}. 
In these analytic star formation theories, supersonic turbulence produces gravitationally unstable density fluctuations and sets the overall fraction of dense star-forming gas. The lognormal volume-weighted density PDF is described by:
\begin{equation}
p_{LN}(s)=\frac{1}{\sqrt{2\pi\sigs^2}}\exp\left(-\frac{(s-s_0)^2}{2\sigs^2}\right) \, ,
\label{eq:LNpdf}
\end{equation}
expressed in terms of the logarithmic density,
\begin{equation} \label{eq:s}
s\equiv\ln{(\rho/\meanrho)} \ ,
\end{equation} 
where $\sigs$ is the standard deviation of the lognormal distribution.
The quantities $\meanrho$ and $\means$ denote the mean density and mean logarithmic density, respectively; the latter of which is related to $\sigs$ by
\begin{equation}
\means=-\frac{1}{2}\,\sigs^2 \ .
\end{equation}
For an isothermal equation of state, the width of the lognormal is determined by the turbulent sonic Mach number \citep{Padoan1997,Federrath2008,Burkhart09,Price11a,Konstandin+2012} and the turbulence driving parameter $b$:
\begin{equation}
\sigs^2=\ln \left[1+b^2 \mach_{s}^{2} \right] \ ,
\label{eqn.sigma}
\end{equation}
where the sonic Mach number depends on the RMS velocity dispersion (v$_{\rm{rms}}$) and the sound speed ($\cs$):
\begin{equation}
\mach_s =\frac{v_{\rm{rms,3D}}}{\cs} \ .
\end{equation}
The turbulent driving parameter, $b$, describes the mix of solenoidal and compressive modes of the turbulent driving and ranges from $b \approx 1/3$ for purely solenoidal driving, to $b \approx 1$ for purely compressive driving \citep{Federrath2008,Federrath10b}. 
Values of $b$ for real clouds occupy the full range from $b\sim0.3$--$1$ \citep[see e.g.,][]{Brunt10a,Price11a,Ginsburg+2013,Kainulainen13b,Federrath+2016b,Menon+2021}, depending on the physical conditions and location of the clouds.
For the simulations used in this work, which are driven with a natural mix of forcing modes, resulting in $b\sim0.4$ \citep{Federrath10b} and $\mach_s = 5$, Eq.~\ref{eqn.sigma} predicts a width of $\sigma_s = 1.27$.

In the presence of magnetic fields, the width of the lognormal distribution is expected to change. 
\citet{PadoanNordlund2011} and \citet{Molina2012} derive an expression for the width of the lognormal that is modified by the ratio of the thermal to magnetic pressure or the plasma $\beta_0$ \citep{Molina2012}:
\begin{equation}
\sigma_{m}^2=\ln \left[1+b^2M_s^2 \left( \frac{\beta_0}{1 + \beta_0} \right) \right] \, .
\label{eq:sigma_m}
\end{equation}
For the simulations used in this work, the plasma $\beta_0 \sim 0.33$, resulting in a predicted value of $\sigma_m = 0.83$. 
In case of a strong guide field, \citet{Beattie+2021} derive an anisotropic version of Eq.~(\ref{eq:sigma_m}). 
In this case, Eq.~(\ref{eq:sigma_m}) still provides a reasonable approximation, although for $\mathcal{M}_A<2$, anisotropies start to play a significant role \citep{Federrath2016b}.

\subsection{A Piecewise Density PDF}\label{sec:piecewisePDF}

\citet{Burkhart2018} proposed using a piecewise lognormal plus power-law density PDF in order to account for the effects of early stellar feedback (e.g., winds, jets, radiation) and self-gravity when calculating the SFE \citep[see also][]{BurkhartMocz2019}. 
In particular, this model accounts for self-gravity by introducing a power-law tail and allows the slope of this power-law, as well as the transition density, to vary in response to the influence of, for example, stellar feedback.
This change was motivated by observational and numerical work suggesting that the density distribution takes the form of a power-law, rather than a lognormal distribution, at the high-density end of the PDF \citep[see e.g.,][]{Klessen2000,Kritsuk+2011,Collins12a,Federrath13a,Girichidis2014,Lombardi2015AA,Leroy2017,Burkhart2017,Chen2018,JaupartChabrier2020,Khullar2021}.

The density PDF proposed by \citet{BurkhartMocz2019} consists of a lognormal distribution at low densities and a power-law distribution at high densities, with a transition point denoted as $s_t=\ln(\rho_t/\rho_0)$), and is described by:
\begin{align}
p_{LN+PL}(s) = 
\begin{cases}
 N\frac{1}{\sqrt{2\pi}\sigs}e^{-\frac{( s - s_0)^2}{2\sigs^2}}  & s < s_t \\
 N C e^{-\alpha s}  & s > s_t \ ,
\end{cases}
\label{eqn.piecewise}
\end{align}
where $\alpha$ is the slope of the power-law tail and $\sigs$ is the width of the lognormal distribution (e.g., given by Eq.~\ref{eqn.sigma}).
The point where the density PDF transitions between the lognormal component and the power-law component can be derived from considerations of continuity and differentiability \citep{Burkhart2017}:
\begin{align}
s_t&=(\alpha -1/2)\sigs^2 \ \ .
\label{eqn.st}
\end{align}
By combining Eq.~\ref{eqn.st} and Eq.~\ref{eqn.sigma}, the transition point can be further related to the Mach number and driving parameter \citep{Burkhart2017}: 
\begin{equation}
s_t=\frac{1}{2}(2\alpha-1)\ln[1+b^2M_s^2] \ \ .
\label{eq:s_t}
\end{equation}

A density PDF with both a lognormal and a power-law component is compatible with a gravo-turbulent model of star formation since the development of a power-law density distribution is consistent with the gas undergoing gravitational contraction while the lognormal distribution still retains some memory of its turbulent initial conditions.
In the initial stages of collapse, the power-law slope will be very steep. 
The slope becomes shallower on a timescale of a freefall time at the critical density \citep{Federrath13a}, and may reach $-1.5$ in the limit of uniform pressure-less spherical collapse \citep{Girichidis2014,JaupartChabrier2020,Khullar2021}. 
Since different values of the power-law slope are expected early in the cloud's evolution, the value of the transition density and fraction of dense self-gravitating gas is also expected to change. 
In addition, gas may cycle through the power-law to lognormal portions of the PDF due to stellar feedback, again adjusting the amount of dense gas available for star formation.

\section{Density PDFs from numerical simulations}
\label{sec:models}

\begin{figure*}[htb!]
\includegraphics[width=\linewidth]{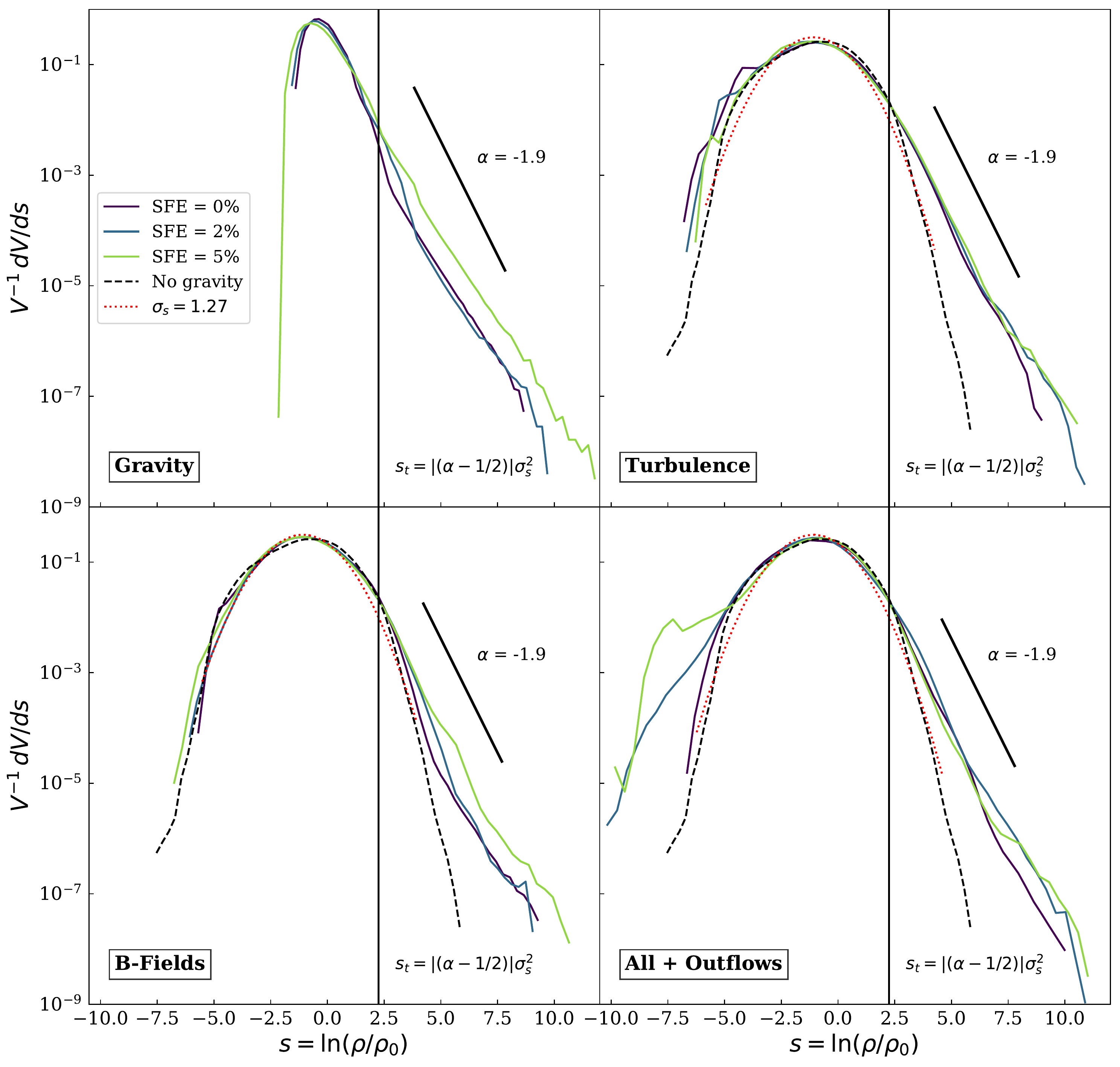}
\caption{Volume-weighted density PDFs for each of the four simulations described in Table~\ref{tab:sims}. Three different snapshots are shown (SFEs~$=0\%$, $2\%$, and $5\%$), where a larger SFE indicates a more advanced time snapshot.  For the \textsc{Turbulence}, \textsc{B-Fields}, and \textsc{All + Outflows} simulations, we overplot two curves for comparison: the dashed black line is a single snapshot from the volume-weighted density PDF for the \textsc{No Gravity} simulation, and the dotted red line is a Gaussian with the theoretically predicted width of $\sigma_s = 1.27$ (as described by Eq.~\ref{eqn.sigma}). A reference line for a power-law tail with a slope of $-1.9$ (the best fit value chosen later in our analysis) is shown as a solid black line. A vertical line indicates the reference transition density ($s_t$ = 2.25) which we use in our subsequent analysis to separate the lognormal and power-law parts of the PDF. 
\label{fig:pdfs}}
\end{figure*}

\begin{figure}[bt]
\includegraphics[width = \linewidth]{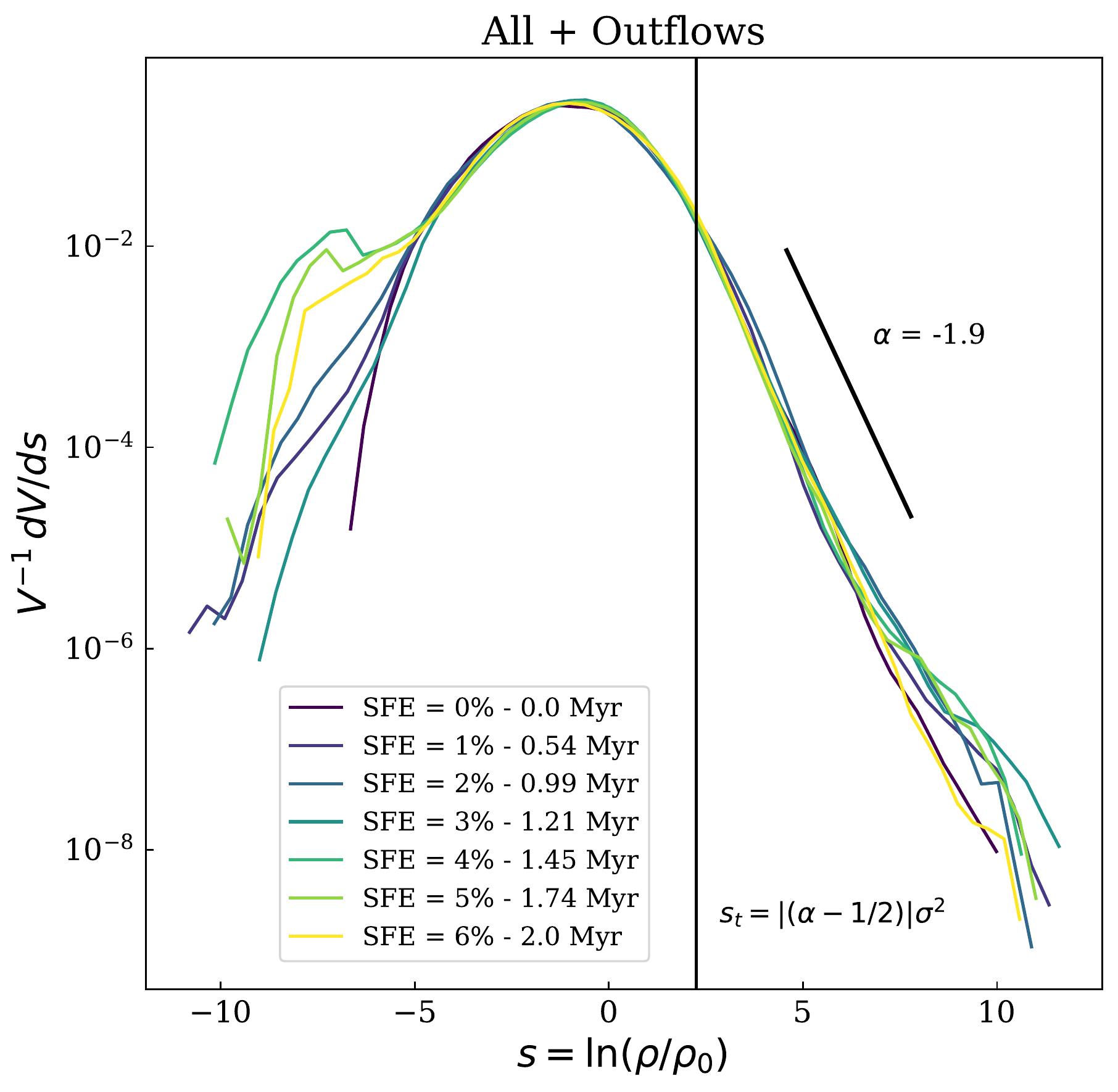}
\caption{Volume-weighted density PDFs for all of the snapshots from the \textsc{All + Outflows} simulation, which includes protostellar outflows and heating. Here, all of the available snapshots are shown to demonstrate the variability of the \textsc{All + Outflows} run. The corresponding times for each snapshot are shown in the legend. Also shown for reference (as a dashed black line) is the volume-weighted density PDF of the \textsc{No Gravity} simulation.  A reference line for a power-law tail with a slope of $-1.9$ is shown as a solid black line. A vertical black line indicates the reference transition density ($s_t = 2.25$).  \label{fig:pdf_feedback}}
\end{figure}

\subsection{Overview of the Density PDFs}

We now consider the shapes of the simulated density PDFs, which each include different physical processes. 
Figure~\ref{fig:pdfs} shows the volume density PDFs for selected snapshots of each of the simulations described in Section~\ref{sec:numerics}.
The SFE~$=0\%$ snapshot is just before the first sink particle forms in each simulation, the SFE~$=2\%$ snapshot is an approximate midpoint in the evolution of each simulation, and the SFE~$=5\%$ snapshot is the latest snapshot where the same SFE was available for all of the simulations.

All simulations that include turbulence (i.e., \textsc{Turbulence}, \textsc{B-Fields}, and \textsc{All + Outflows}) show a prominent lognormal peak in the PDF, in agreement with expectations for supersonic turbulence (see Section~\ref{sec:LNmodels}). 
For reference, we overplot a single snapshot of the density PDF for the \textsc{No Gravity} run (dashed black line), which includes only turbulence and magnetic fields but no gravity, and a theoretical lognormal shape with a width of $\sigma_s = 1.27$ (red dotted line), as predicted by Eq.~\ref{eqn.sigma}, using the driving parameters of the simulations: a sonic Mach number of 5 and a driving parameter $b=0.4$. 
Both lines match the lognormal peak in all three simulations reasonably well. 
The minor discrepancy with the theoretical curve is likely due to statistical variations of the PDF that occur between individual snapshots \citep[see, e.g.,][]{Federrath2008}. 
We do not make this comparison for the \textsc{Gravity} simulation, as it lacks any turbulent driving and the narrow lognormal peak of its PDF is simply an imprint of the Gaussian random density field used to initialize this simulation.

At high densities, the simulated PDFs quickly depart from a pure lognormal shape as a power-law tail forms due to the influence of self-gravity, in agreement with the theoretical expectation discussed earlier. 
The power-law tails have slopes close to $-1.9$ on average, although there is also some evidence of multiple power-law tails with different slopes. 
Two or more power-law slopes have previously been observed in molecular clouds, for example in \cite{Schneider+2015}. 
This may be because of the time-dependence of the emergence of different power-law slopes due to gravitational collapse \citep{Burkhart2017,JaupartChabrier2020} and/or due to the formation of rotationally supported structures, i.e., formation of accretion disks \citep[][]{Khullar2021}. 
The latter may only be seen in the simulation with outflows as this simulation has sufficient resolution to capture hints of accretion disk formation.

To gauge the transition point between the lognormal peak and power-law tail, we overplot the reference transition density of $s_{t} = 2.25$ (vertical black line) calculated from Eq.~\ref{eqn.st} and Eq.~\ref{eqn.sigma}, using a power-law tail slope of $-1.9$ and the lognormal width of $\sigma_s = 1.27$. 
As the comparison with the \textsc{No Gravity} run shows, this value of $\s_t$ is close to the density at which the power-law tail forms and the PDF deviates from a lognormal shape. 
Note that in the \textsc{Gravity} simulation, which lacks any turbulent driving, we do not expect a lognormal distribution to form and the power-law tail forms at a significantly lower density.
Given that our subsequent analysis mainly focuses on the effects of magnetic fields and protostellar feedback, we choose to use the same $s_t$ in the \textsc{Gravity} simulation as in the runs with turbulence driving. 
The reference transition density is shown in the \textsc{Gravity} panel of Fig.~\ref{fig:pdfs} for the sake of comparison with the other panels.

Finally, in addition to the high-density power-law tail, the \textsc{All + Outflows} simulation also deviates from a lognormal distribution at the \textit{low-density end of the PDF.}  
When protostellar outflows and heating are included, the low-density end of the PDF exhibits significant fluctuations as the simulation progresses. 
Figure~\ref{fig:pdf_feedback} shows all available snapshots of the \textsc{All + Outflows} simulation, demonstrating that the significant fluctuations in the low-density tail of the PDF are time-varying.

Deviations from lognormality at both low and high densities for each of the runs are discussed further below in Section~\ref{sec:fits}.

\subsection{Fits to the Density PDFs} \label{sec:fits}

\begin{figure}[htb]
\includegraphics[width = 0.9 \linewidth]{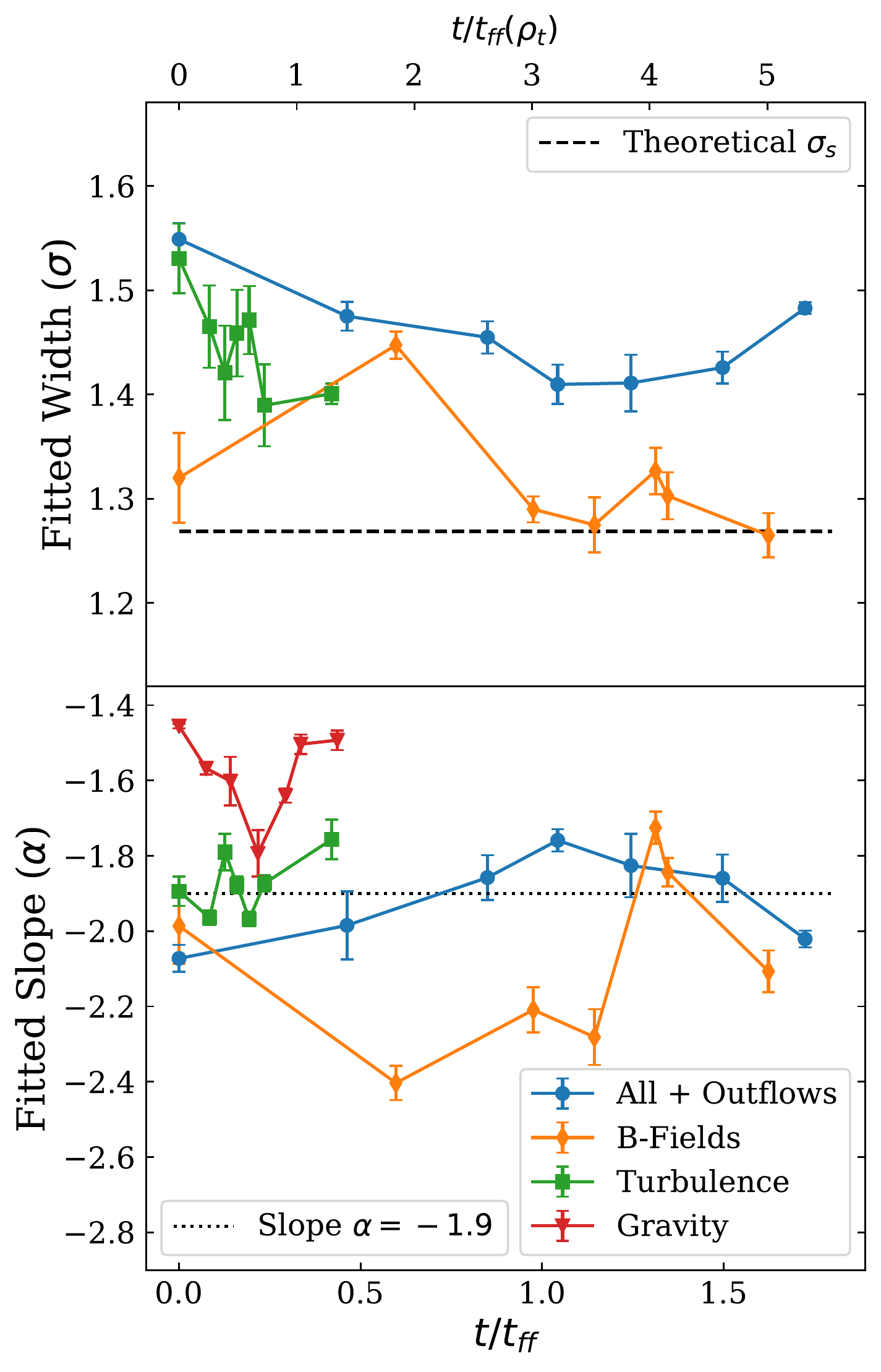}
\caption{\textbf{Top:} The width of the fitted Gaussian curves for each of the snapshots for all of the simulations. The error bars are the 1-$\sigma$ uncertainties from the covariance matrix of each fit. Also shown is the theoretical value for the width of the Gaussian peak due to only turbulence, $\sigma_s = 1.27$, as given by Eq.~\ref{eqn.sigma} if we assume a fixed $b=0.4$. 
\textbf{Bottom:} The slope of the fitted power-law tail for each snapshot of each simulation.
Also shown is a constant slope of $\alpha = -1.9$ that we use as a reference in Figs.~\ref{fig:pdfs} and \ref{fig:pdf_feedback} (dotted black line).
\textbf{Both:} Each panel is shown as a function of time, where time is measured in freefall times of the average density ($t_{\text{ff}} \approx 1.16$ Myr) on the lower axis and in freefall times of the reference transition density on the upper axis ($t_{\text{ff}}(\rho_{t}) \approx 0.38$ Myr). Time is measured since the $0\%$ snapshot, i.e., right before the first sink particle is formed. 
\label{fig:widths}}
\end{figure}

\begin{figure}[htb]
\includegraphics[width = 0.9 \linewidth]{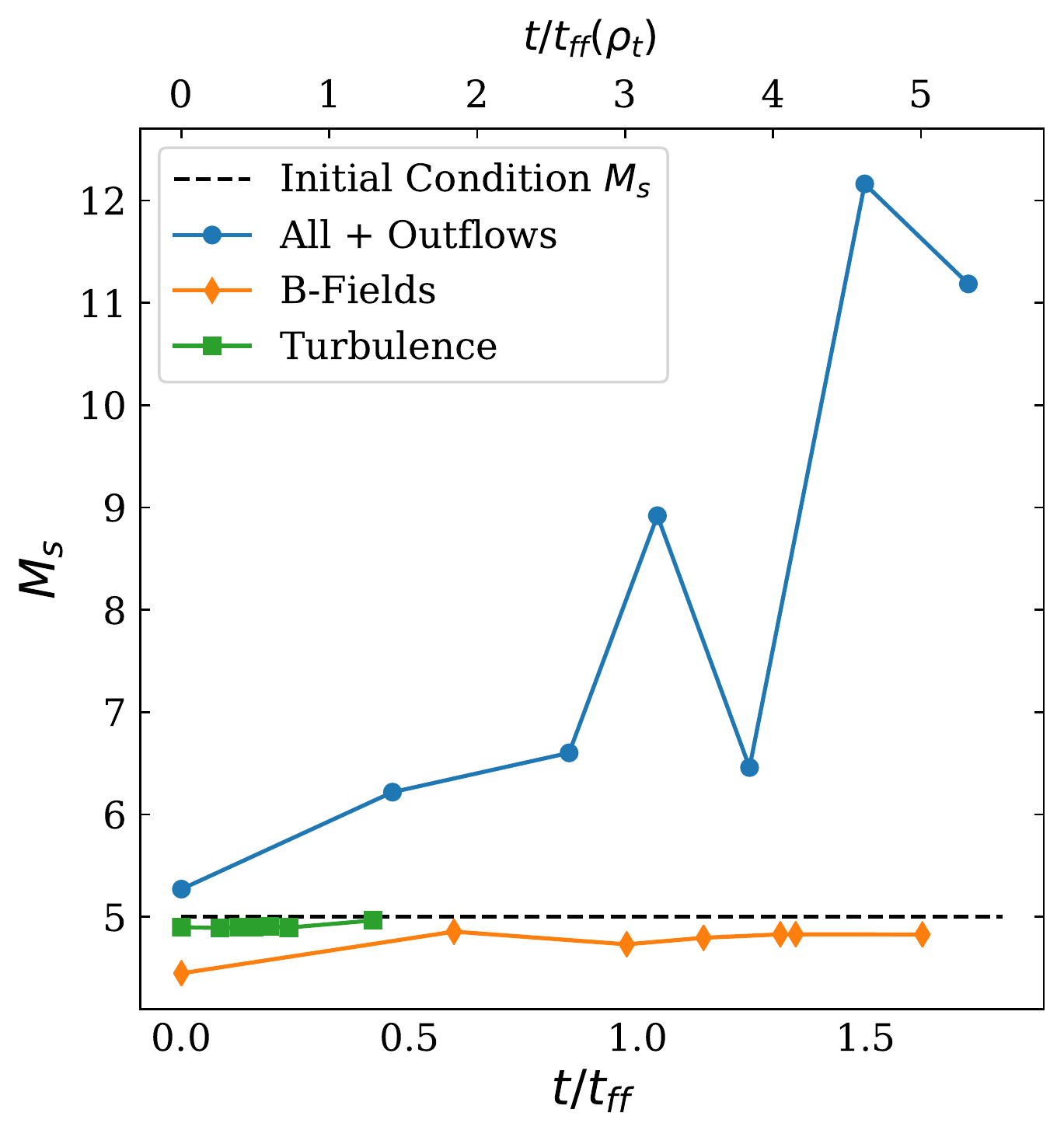}
\caption{The RMS sonic Mach number ($\mach_{s}$) of each simulation as a function of time and the initial conditions Mach number ($\mach_{s} = 5$; dashed black line). The x-axes are the same as in Fig.~\ref{fig:widths}.
\label{fig:mach}}
\end{figure}

\begin{figure}[htb]
\includegraphics[width = 0.9 \linewidth]{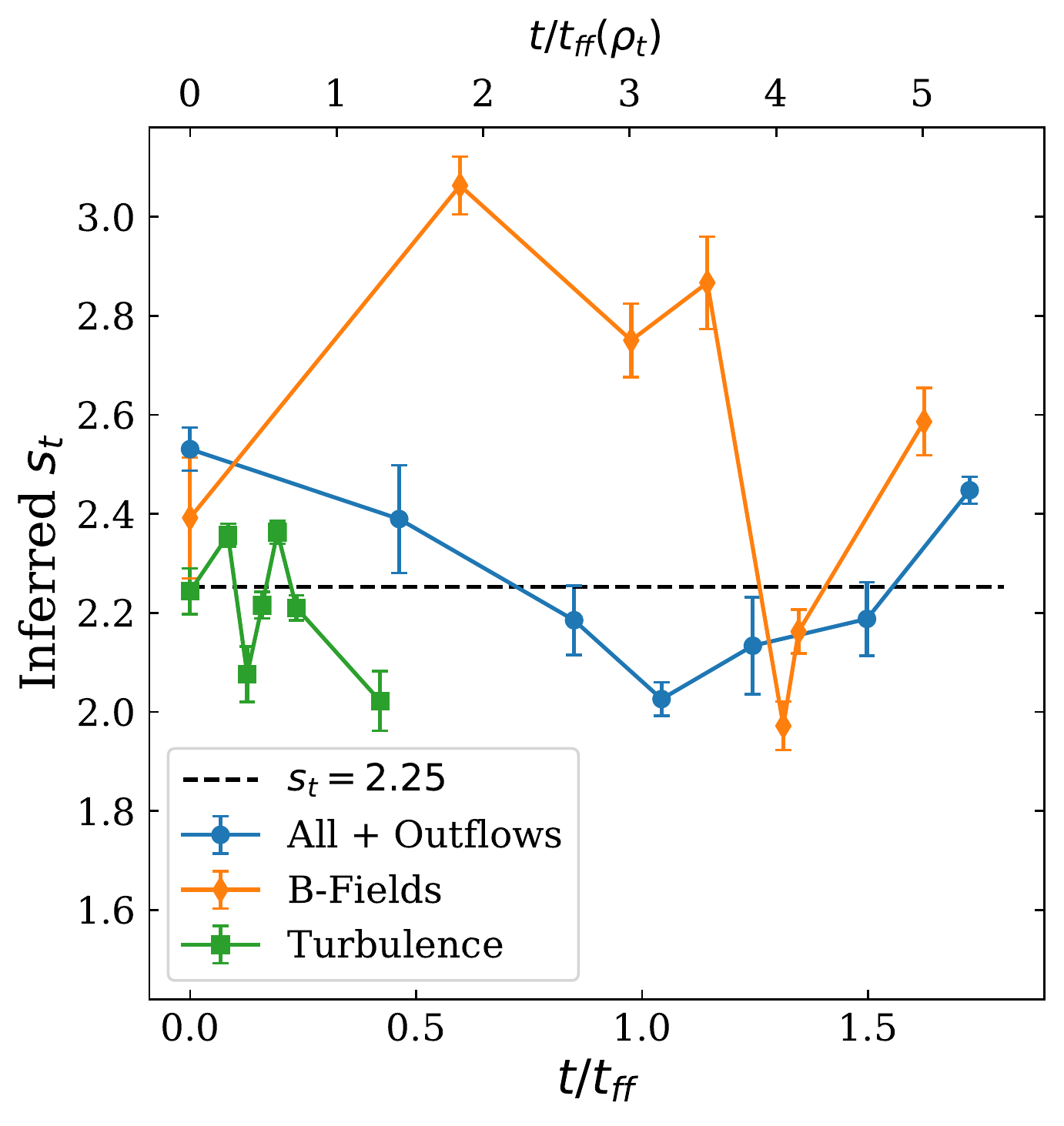}
\caption{The inferred transition densities for each snapshot, calculated using Eq.~\ref{eqn.st} and the fitted power-law slopes (shown in Fig.~\ref{fig:widths}), but using the fixed value of $\sigma_s = 1.27$. The errorbars are propagated forward from the errors on the fitted slopes.
The dashed line shows the reference transition density ($s_t=2.25$) corresponding to a fixed power-law slope or $-1.9$ and the theoretical value of $\sigma_s = 1.27$.
The x-axes are the same as in Fig.~\ref{fig:widths}. 
\label{fig:st_vs_time}}
\end{figure}

To compare our density PDFs with the models discussed in Section~\ref{sec:pdfs}, we fit a lognormal curve to the peak of the density PDF and a power-law relation to the high-density tail of the PDF. 
We use a separate lognormal fit and power-law fit so that we can focus separately on the behavior of the PDF in the lognormal portion of the PDF and in the power-law portion. 
Recent work has begun to implement a piecewise lognormal plus power-law fit \citep{Khullar2021}, but for this work we choose to use separate fits in accordance with the approach used in, e.g., \cite{Kritsuk+2011,Schneider+2015,schneider15,Alves2017AA,Soler2019}.

We fit a Gaussian curve (i.e., a parabola in log space) to the peak of the density PDF using \verb|curve_fit| from the \verb|scipy optimize| package, which performs a non-linear least squares fit to the data.
For each snapshot, we performed a single Gaussian fit to a specific range of densities selected to exclude regions of the PDF that appeared during visual inspection to deviate significantly from a lognormal distribution. 
The upper limit for all snapshots was set at $s=2.0$ in order to exclude regions of the PDF that visually appear to not be purely lognormal.  
For the \textsc{Gravity}, \textsc{Turbulence}, and \textsc{B-Fields} simulations no lower limit was set and all bins below $s=2.0$ were used for the fits.
For the \textsc{All + Outflows} simulation, we use a lower limit of $s=-4.0$ in order to exclude the excess of low-density gas in the PDF, produced by protostellar outflows, that deviates significantly from a lognormal PDF (see, for example, Fig.~\ref{fig:pdf_feedback}).
These upper and lower limits ensure that each fit minimized the influence of deviations from lognormality at high and low densities.
The \verb|curve_fit| method produces a covariance matrix, from which we derived 1-$\sigma$ uncertainty errorbars for each fit.

We also used \verb|curve_fit| to fit a power-law function at the high-density ends of the PDFs. 
For each snapshot, we performed a single linear fit ($\log({\rm PDF}) = \alpha \, s \, + \, \beta$, in log space, where $\alpha$ is the slope and $\beta$ is the $y$-intercept), with a fitting range between $s=3.0 -8.0$.
We selected the upper limit of the fitting range to capture the behavior of the high-density end of the PDF while minimizing the impact of the PDF fluctuations at high densities.
For the SFE $= 1 \%$ snapshot of the \textsc{B-Fields} simulation, we adjusted the fitting range to be between $s=3.0- 6.0$ because the PDF for this snapshot drops off dramatically above $s=6.0$ and skews the fit to a much steeper slope.
This is likely due to the formation of a single $3.9$ M$_{\odot}$ sink particle just before the time of the snapshot.
For all of the snapshots, we selected $s=3.0$ as the lower limit of the fitting range to exclude portions of the PDFs that are clearly no longer purely linear under visual inspection.
We derived the 1-$\sigma$ uncertainties for each fit from the covariance matrix produced by the \verb|curve_fit| procedure.

The widths and slopes of the fits described above are shown in Fig.~\ref{fig:widths}. 
The top panel of Fig.~\ref{fig:widths} shows the widths of the Gaussian fits for all of the snapshots, with the error bars derived from the covariance matrix, for all runs except the \textsc{Gravity} simulation.  
The \textsc{Gravity} simulation is not shown because it does not include turbulence and the narrow lognormal peak apparent in the \textsc{Gravity} panel of Fig.~\ref{fig:pdfs} is due to the initial Gaussian random field.
The bottom panel of Fig.~\ref{fig:widths} shows the fitted slopes for each snapshot of each simulation, with the error bars showing the 1-$\sigma$ uncertainty of the fit.
We find that a slope of $\alpha = -1.9$ (shown with the dotted line) is a reasonable value for the power-law slopes, although we do see values ranging from $\alpha = -1.4$ to $\alpha = -2.4$. 
Thus, we will use the value of $\alpha = -1.9$ for determining a constant transition density in our later analysis. 
This value of the slope is also shown in Figs.~\ref{fig:pdfs}~or~\ref{fig:pdf_feedback} for reference.

Interestingly, the upper panel of Fig.~\ref{fig:widths} shows that the fitted widths fall above the predicted value of $\sigma_s=1.27$ that we find using Eq.~\ref{eqn.sigma} if we assume that $\mach_s = 5$ and $b = 0.4$ are fixed (black dashed line in the figure). 
However, when physics beyond turbulence is included, the values of both $\mach_s$ and $b$ can change \citep{JaupartChabrier2020,Khullar2021}. 
For example, gravity will possibly drive more compressive modes of turbulence \citep{JaupartChabrier2020,Khullar2021}. 
The  formation of the power-law tail at high densities may also bias the fits of the lognormal portion of the density PDF towards a wider value. 
Likewise, outflows may induce more solenoidal and compressive modes of turbulence \citep[see also][]{Offner+2017,RosenKrumholz2020}.

We check whether an increase in the sonic Mach number may be the reason for this increase in the fitted PDF widths by plotting the evolution of the sonic Mach numbers vs. time in Fig.~\ref{fig:mach}. 
The solid lines show the RMS sonic Mach number for each snapshot as calculated using the magnitude of the gas velocity and the sound speed.
The dashed black line shows the sonic Mach number that is continuously driven during the simulation run ($\mach_s = 5$).

The RMS sonic Mach numbers for the \textsc{Turbulence} and \textsc{B-Fields} simulations remain fairly close to the input value.
In contrast, the RMS sonic Mach number for the \textsc{All + Outflows} simulation rises significantly over time, due to the added kinetic energy from the protostellar outflows.
Because the RMS sonic Mach numbers do not diverge from the driving value of $\mach_{s} = 5$ with the inclusion of magnetic fields and gravity, the fitted widths in Fig.~\ref{fig:widths} for those simulations cannot be explained solely by a change in sonic Mach number. 
In the case of the \textsc{All + Outflows} simulation, a sonic Mach number of $\approx 7$ could explain the fitted widths. 
However, at later snapshots the RMS Mach number increases dramatically, whereas the fitted width does not. 
Furthermore, at early snapshots the RMS Mach number is too low to reflect the PDF width.
These results suggest that either $b$ must be changing or the fitting procedure overestimates the width due to the influence of the power-law tail and/or due to the low-density fluctuations from the outflows.

\subsection{Transition Density Between the Lognormal and the Power-law}

Using our fitted slopes and the width predicted by Eq.~\ref{eqn.sigma}, we calculate the transition density (Eq.~\ref{eqn.st}) for each snapshot. 
As discussed above, the fitted widths (shown in Fig.~\ref{fig:widths}) are wider than the value predicted by Eq.~\ref{eqn.st}. 
This may be due to a bias in the fits from the influence of the power-law tail or due to a changing value of $b$ from additional compressive motions. 
Furthermore, in \cite{Burkhart2018}, the width used to calculate the transition density is determined only by the properties of the large-scale turbulent flow (i.e, the driving-scale sonic Mach number and the turbulent driving parameter $b$). 
In the context of these simulations, this suggests using the known parameters of the driven turbulence, $\mach_{s}=5$ and  the driving-scale $b=0.4$, to infer the width. 
Thus, we calculate the transition density using the width predicted by Eq.~\ref{eqn.sigma}, which depends on only the large-scale, driven turbulence, in accordance with \cite{Burkhart2018}. 

We plot the computed values of $s_t$ vs. time in Fig.~\ref{fig:st_vs_time}.
We show a reference transition density of $s_{t} = 2.25$ (dashed black line), computed using a reference power-law slope of $\alpha = -1.9$ and the width predicted by Eq.~\ref{eqn.sigma}. 
This reference transition density $s_t=2.25$ is close to a median value around which the fluctuations occur.

There is significant variation of $s_t$ with time, which reflects the variation in the fitted slopes, as shown in the lower panel of Fig.~\ref{fig:widths}.
The inferred values of $s_t$ for the \textsc{Turbulence} and \textsc{All + Outflows} simulations generally correspond to the reference value across the snapshots. 
The \textsc{B-Fields} simulation shows significantly more variation of $s_t$ with time. 
This is because the power-law slope is significantly more time variable, for the reasons discussed above.

We overplot the reference transition density ($s_t = 2.25$) as a vertical line in Figs.~\ref{fig:pdfs}~and~\ref{fig:pdf_feedback}.
The $s_t = 2.25$ transition density provides a compelling point of divergence between the lognormal and power-law portions of the PDF. 
It closely matches the density at which the simulation without gravity (black dashed lines) and the simulations with gravity and driven turbulence no longer track each other. 
We therefore use the reference transition density of $s_t = 2.25$ for all of our subsequent calculations (Section~\ref{sec:mass_evol}) in order to provide a single reference density for comparison of all simulations.
Future work will further investigate a variable transition density and different methods for fitting the transition density.

\subsection{Mass Flow from Diffuse to Collapsing Gas As Traced by the PDF}\label{sec:mass_evol}

\begin{figure*}[ht!]
\includegraphics[width = \linewidth]{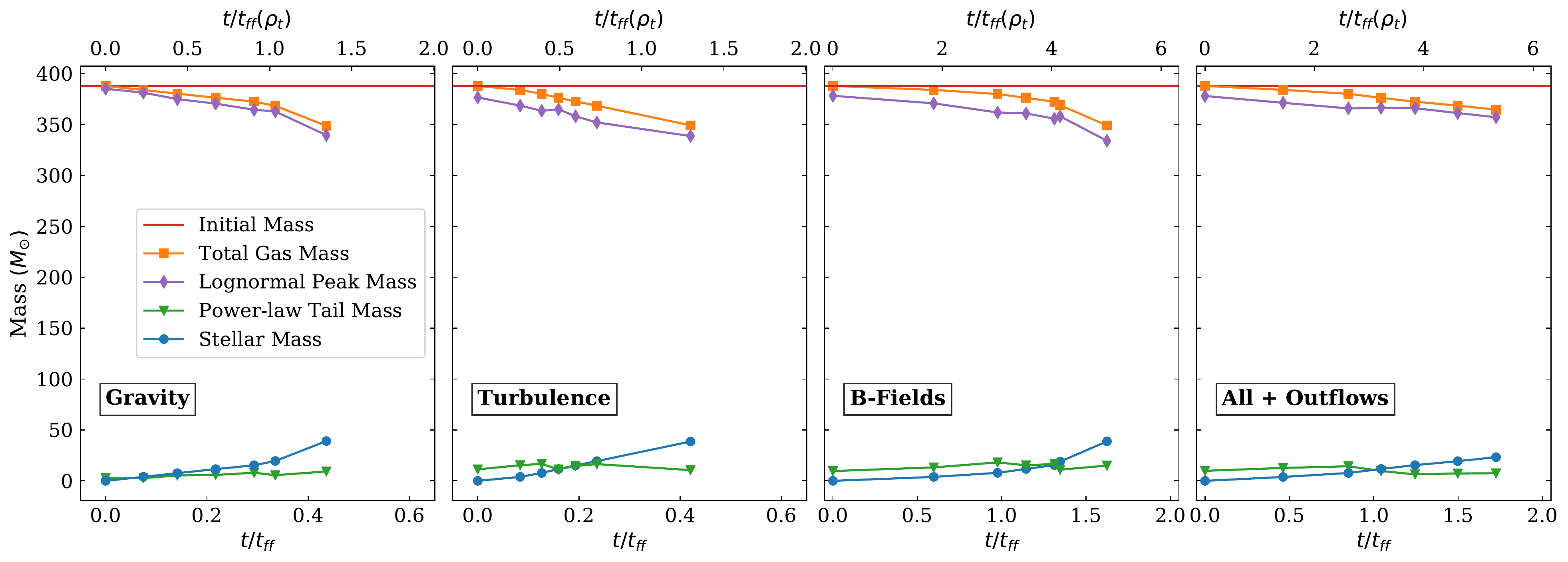}
\caption{The total mass contained in each component of the density PDF. Initial Mass refers to the initial total gas mass of the simulation and is equal to $388 \, M_{\odot}$. Total Gas Mass refers to the total mass contained within the density PDF. Lognormal Mass is the mass contained below the reference transition density ($s < s_t = 2.25$). Power-law Tail Mass refers to any gas above the reference transition density ($s > s_t = 2.25$). Stellar Mass (the blue points) refers to the mass contained within the sink particles. Each panel is shown as a function of time, where the same x-axes are used as in Fig.~\ref{fig:widths}. Note the  difference in the x-axes ranges for the \textsc{Gravity} and \textsc{Turbulence} simulations. \label{fig:massvtime}} 
\end{figure*}

In this section we analyze the total mass contained in the lognormal and power-law portions of the density PDF and study how that mass moves into the sink particles.
We considered the total mass contained in the sink particles, in the density PDF overall, in the power-law tail, and in the lognormal part of the PDF.
The power-law tail is defined as any gas that is denser than the reference transition density, $s_t = 2.25$.
The lognormal peak is defined as any gas that is less dense than the reference transition density.

Figure~\ref{fig:massvtime} shows the mass contained within each component of the density PDF as a function of time for each simulation.
The horizontal red line shows the initial mass of the simulation for reference. The total gas mass (orange line) decreases in time as sink particles form and accrete mass (blue line).
Figure~\ref{fig:massvtime} also separately shows the mass of the diffuse gas in the lognormal portion of the PDF (i.e., at $s < s_t$; purple line) and the dense, self-gravitating mass in the power-law tail (i.e., at $s > s_t$; green line).
The power-law tail mass is generally stable in time for all of the simulations, while the lognormal peak mass steadily decreases in step with the total gas mass decrease. 
Overall, the \textsc{Gravity} and \textsc{Turbulence} simulations exchange gas mass into stellar mass on a significantly faster time scale than the other two simulations. 
Because these simulations evolve much faster, the x-axes for these two simulations are different than the other panels in order to make the evolution easier to see.

Figure~\ref{fig:massvtime} shows that the diffuse gas reservoir is replenishing the power-law tail mass at a rate that is equal to the star formation rate, such that it maintains a stable power-law tail. 
The main difference between the four simulations is that the inclusion of magnetic fields and outflow feedback results in slower accretion of gas from the lognormal portion of the PDF to the power-law tail, in agreement with previous work \citep[see e.g.,][]{Cunningham12a,Federrath2015,Offner+2017,RosenKrumholz2020}. 

\begin{figure*}[ht!]
\centering
\includegraphics[width = 0.8\linewidth]{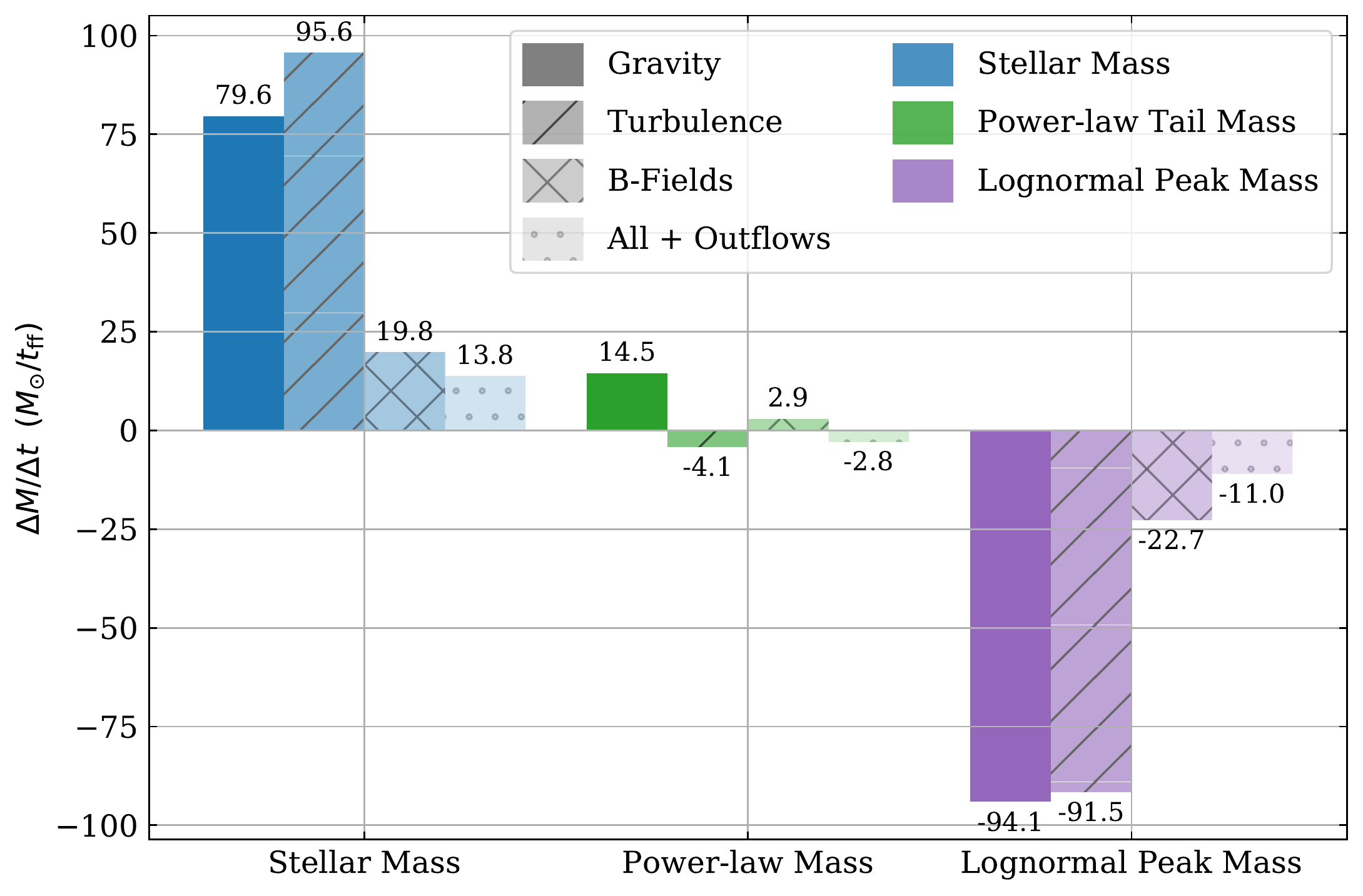}
\caption{The slopes of linear fits to all points for each of the PDF components in Fig.~\ref{fig:massvtime}, which indicates the rate of change in mass of the PDF components. The hatching and saturation of each bar indicate the simulation, while the color indicates the mass component (also indicated on the x-axis). For all four simulations, the power-law mass has the smallest slope, suggesting it changes the least for all of the simulations. The numerical value of each slope (in $M_{\odot}/t_{\rm ff}$) is also shown. \label{fig:slopes}}  
\end{figure*}

\begin{figure*}[ht!]
\centering
\includegraphics[width = \linewidth]{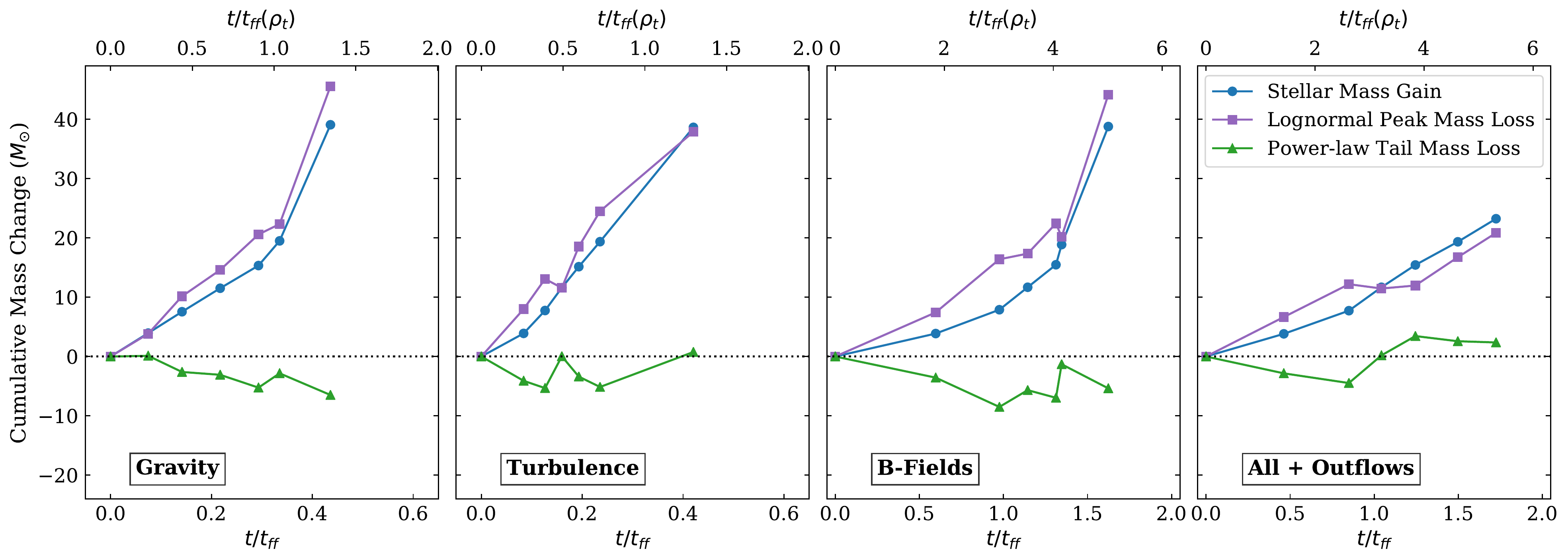}
\caption{The cumulative change in mass for each component of the density PDF as a function of time. Note that the total stellar mass is shown as a positive mass gain, while the power-law tail mass and the logormal peak mass are inverted to show the cumulative mass lost from the corresponding region of the density PDF. This inversion highlights the fact that any mass lost from the lognormal peak or the power-law tail mass has a corresponding increase in another component of the PDF. Each panel is shown as a function of time, where the same x-axes are used as in Fig.~\ref{fig:widths}. Again, note the  difference in the x-axes ranges for the \textsc{Gravity} and \textsc{Turbulence} simulations.  \label{fig:mass_loss}}  
\end{figure*}

To illustrate this difference in the accretion rates, Fig.~\ref{fig:slopes} compares the time-averaged rates of change for each of the PDF components. 
To produce this plot, we fit a linear function to each of the trends in Fig.~\ref{fig:massvtime} using the \verb|LinearRegression| method from \verb|scikit-learn|'s \verb|linear_model|, which performs an ordinary least squares linear regression. 
Figure~\ref{fig:slopes} shows the slopes, or $\Delta M/\Delta t$, of each of these linear regression fits. 
The total gas mass $\Delta M/\Delta t$ values are not shown in Fig.~\ref{fig:slopes} because they are identical and opposite to the stellar mass slopes.
Note when considering Fig.~\ref{fig:slopes} that the linear regressions fit the overall trend and do not capture variations in the slope. 
For example, a fit to only the final two or three points of the \textsc{Gravity} case would result in a much steeper slope than is measured when fitting all of the points. 

For all four simulations, the change in the power-law tail mass with time is small, as expected from Fig.~\ref{fig:massvtime}. 
We also find that the slopes for the lognormal and stellar mass components are similar to each other in all four cases.
However, for the \textsc{B-Fields} and \textsc{All + Outflows} simulations, the change in the lognormal mass with time is comparatively much smaller than the other simulations.

We also consider the cumulative change in mass for the power-law tail mass, the lognormal peak mass, and the stellar mass of each simulation in Fig.~\ref{fig:mass_loss}.
We plot the mass \emph{lost} from the power-law tail (green line) and the lognormal peak (purple line), as well as the mass \emph{gained} in the sinks (blue line).  
In other words, Fig.~\ref{fig:mass_loss} shows the negative, cumulative change in mass for the power-law tail and the lognormal peak but the positive, cumulative change in the sink mass.
This allows us to track which part of the density PDF is ultimately contributing to the growth of the sink particles.

We see a similar situation in all four simulations. 
The lognormal mass loss and the total stellar mass gain track each other closely, while the power-law tail mass loss remains close to zero or slightly negative (indicating a small mass gain). 
This suggests, in agreement with the discussion of Figs.~\ref{fig:massvtime}~and~\ref{fig:slopes}, that the mass in the power-law portion of the PDF is replenished from the lognormal portion at the same rate that it loses mass to the sink particles. 
However, the time scale on which this occurs is different for each simulation. 
As seen above, the two simulations with magnetic fields (\textsc{B-Fields} and \textsc{All + Outflows}) evolve much more slowly. 
Similar to Fig.~\ref{fig:massvtime}, the x-axes are different between the runs with and without magnetic fields. 

Furthermore, for the \textsc{All + Outflows} simulation, the cumulative mass change remains shallower at all time steps than in the other simulations, suggesting a slower star formation rate for this simulation.
The lognormal peak mass loss drops slightly below the stellar mass gain in the last few time steps of the \textsc{All + Outflows} simulation and the power-law tail mass loss becomes slightly positive.
This suggests that, in these time steps, protostellar outflows slow down the rate at which the lognormal replenishes the power-law tail and the power-law tail is slightly depleted.
This may be due to the additional kinetic energy injected by protostellar outflows into nearby gas, making it difficult for diffuse gas to collapse.

Interestingly, the \textsc{Turbulence} simulation converts all of the power-law tail mass from $t=0$ ($M_{\rm PL}=11.3$ $M_\sun$ at $t=0$) into stars within a free-fall time of the transition density, i.e., by $t=t_{\rm ff}(\rho_{\rm t})$. 
This suggests that large-scale driven turbulence does not provide any additional support in the collapsing dense gas.
This agrees with the fact that turbulence can lead to increased fragmentation as well as longer overall collapse timescales due to the production of additional diffuse gas (i.e., density fluctuations via the creation of a lognormal distribution). 
Thus, turbulence also slows the global rate of mass transfer from the lognormal to the power-law tail  \citep{Federrath2015,Offner09b, Rosen+2019} on time scales longer than $t=t_{\rm ff}(\rho_0)$. 

When magnetic fields  are included, all of the mass rates of change are smaller, indicating that the mass transfer from the diffuse, lognormal portion of the PDF to the dense gas and into the stars is stalled and that the depletion times in these simulations are significantly longer than the \textsc{Gravity} and \textsc{Turbulence} runs. 
Magnetic fields are able to provide support to the gas against collapse and converging motions at all scales and densities.
This is in agreement with the findings of \cite{RosenKrumholz2020}, who find that magnetic fields strongly suppress fragmentation.
In addition, the mass rates of change are even smaller for the \textsc{All + Outflows} case. 
This effect may be due in part to mass lost to the outflows or may be a result of the increased kinetic energy seen earlier. 
Our results suggest that both magnetic fields and protostellar outflows are extremely important for inhibiting the accretion of diffuse gas into collapsing gas and of collapsing gas into stars, and that turbulence alone does very little to prevent collapse in the dense gas of the power-law tail.

\section{Discussion} \label{sec:discussion}

In this paper, we investigated the density PDFs of a suite of four simulations of a 2~pc region of a star-forming molecular cloud.
We find that the inclusion of physical processes beyond turbulence results in deviations from a pure lognormal density PDF.
Specifically, self-gravity produces a power-law tail in the high-density, collapsing gas, and protostellar feedback produces other deviations from lognormallity, particularly at the low-density end of the PDF.
Thus, the inclusion of self-gravity and protostellar feedback in realistic cloud environments implies non-lognormal density PDFs.

Our results agree with recent observed density PDFs which have suggested evidence of non-lognormality.
For example, some recent observational and numerical work shows that the density PDF of the dense gas in molecular clouds is found to have a power-law distribution (for observational work see e.g., \citealt{Collins12a,MyersP2015,schneider15,Alves2017AA,MyersP2017,Kainulainen2017,Chen2018,Ma+2021}; for numerical and theoretical work, see e.g., \citealt{Klessen2000,Kritsuk+2011,Federrath13a,Girichidis2014,Lombardi2015AA,Burkhart2017,Mocz2017,Chen2017,JaupartChabrier2020,Khullar2021}).
However, with the inclusion of stellar feedback, we no longer observe a lognormal density PDF at the lowest densities and instead observe significant time-dependent fluctuations in the low-density part of the PDF. 
We only see evidence of a stable lognormal distribution near the mean density.
Future work with these simulations will include constructing the column density PDFs and comparing them to observations such as those in \cite{Lombardi2015AA}.

We also find that the measured lognormal widths from these simulations does not match the width predicted by the variance-sonic Mach number relationship (Eq.~\ref{eqn.sigma}), if we assume that the large-scale turbulence driving parameter $b=0.4$ and the sonic Mach number $\mach_s = 5$ are fixed throughout the simulations (e.g., Fig. 5, top panel). 
We find that the RMS sonic Mach number cannot account for the wider PDFs, although we do find that outflows inject additional kinetic energy and increase the value of $\mach_s$ (see Fig.~\ref{fig:mach}), in agreement with the findings of \cite{Rosen+2020}. 
Our PDF fit may be biased to wider values due to the influence of the power-law tail. 
It is also likely that gravity changes the turbulent driving parameter, $b$, by increasing the production of compressive modes \citep[e.g.,][]{JaupartChabrier2020,Khullar2021}. 
However, \cite{Pan+2016} find in large-scale ($\sim200\mathrm{pc}$) simulations that $b$ stays relatively low, though their setup differs significantly from ours in that gravitational collapse on small scales is not well resolved; hence, there is only relatively little effect from gravity on the density distributions in their simulations. 
Further work is needed to understand whether and how the turbulent driving parameter $b$ changes with the inclusion of physics beyond supersonic turbulence.
In summary, we find that the inclusion of physics beyond turbulence induces significant changes in the density PDF (the power-law tail and the low-density fluctuations discussed above).

Our results point to the importance of the overall rate at which mass flows from diffuse to dense states in setting the star formation rate.  In these simulations, the amount of dense gas in the power-law is similar and remains roughly constant in time (i.e., Figs.~\ref{fig:massvtime}~and~\ref{fig:slopes}).  
Figure~\ref{fig:mass_loss} shows that mass moves from the lognormal to the power-law tail at the same rate that mass moves into sink particles, resulting in a minimal change in the total mass contained in the power-law tail. This is true for all four simulations studied here, despite the different physics included.

We find that turbulence does not provide significant supportive pressure in the power-law tail. 
Indeed, the presence of shocks in a supersonic flow will enhance collapse locally around the shock. 
Furthermore, turbulence decays towards smaller scales, corresponding to less support for higher density gas in collapsing environments. 
However, shocks also produce rarefied gas and hence a broad lognormal distribution, which helps to slow collapse in the diffuse gas, as the less dense gas has a longer free-fall time. 
Thus, over a free-fall time, turbulence overall acts to reduce the star formation rate.

Our results provide evidence for the importance of mass cycling on cloud scales in setting lower SFRs when magnetic fields and stellar feedback are considered.  
The time varying fluctuations in the low-density end of the volume density PDF for the \textsc{All + Outflows} simulation provide evidence of mass cycling between high and low densities.
The simulated clouds collapse under the influence of gravity, which leads to the local accretion of mass onto the sink particles (e.g., Figs.~\ref{fig:massvtime},~\ref{fig:slopes},~and~\ref{fig:mass_loss}).
However, Fig.~\ref{fig:pdf_feedback} demonstrates that the amount of low-density gas also changes with time, initially growing rapidly and then falling again.
This suggests that gas is cycling between high- and low-density states.
\cite{Semenov2017,Semenov2018} show evidence of a similar process on galaxy scales and provide a framework for connecting gas depletion times and star formation rates with the timescale of this cycling.
Quantifying the gas cycling in our turbulent box simulations will be the focus of a separate study.
Future work will take a closer look at how  mass flow occurs, how it connects to analytic star formation models that use the PDF, and how this process depends on a wider parameter spread of magnetic field strength and turbulent driving.

\section{Conclusions}\label{sec:conclusion}

In this paper, we analyze a suite of four hydrodynamical simulations in order to gain a deeper understanding of how individual components of the density PDF are affected by turbulence, magnetic fields, gravity, and protostellar outflow feedback.
We find that:

\begin{itemize}

    \item The inclusion of protostellar outflows produces time-varying, non-lognormal features in the low-density end of the density PDF, which results in an excess of diffuse gas relative to simulations that do not include feedback. 
    
    \item The inclusion of self-gravity produces a power-law tail at the high-density end of the density PDF. The power-law tail mass and slope does not significantly vary in time despite the presence of outflow feedback. 
    
    \item The density distribution retains a lognormal shape only around the mean density, reflecting the presence of large-scale turbulence. However, we measure wider lognormal fits than would be predicted by the variance-sonic Mach number relationship using the driven Mach number and the driving-scale forcing parameter $b$. This may be because the individual lognormal fits are biased by the growth of the power-law tail and/or because gravity modifies the value of $b$. 
    
    \item Mass flows from the lognormal portion of the density PDF to the power-law tail at a rate that is nearly equal to the sink particle mass accretion. This implies that stars form at the rate at which the diffuse gas contracts and replenishes the power-law tail of the density PDF.  
    
    \item Unlike magnetic fields, we find that driven turbulence does not provide significant support in the dense gas of the power-law portion of the PDF. Over a free-fall time at the transition density, our supersonic hydrodynamic turbulence simulation converts all the mass in the power-law tail into stars. 
    
    \item The inclusion of stellar feedback and magnetic fields significantly slows the transference of mass from gas into stars, as well as from the diffuse gas to the dense collapsing gas. 
    
    \item Our results shed light on the way that the gas mass flows between different parts of the density PDF and highlights the importance of such flows in setting slow star formation rates in molecular clouds. Low star formation rates result from the slowdown of the net mass transfer due to the combined effect of turbulence, gravity, magnetic fields, and protostellar feedback, in line with the picture of slow star formation due to gas cycling between dense and diffuse states \citep{Semenov2017}.
    
\end{itemize}

\acknowledgments

S.A.~thanks Matthew E.~Orr and Charlotte Olsen for their valuable comments on the manuscript. The analysis and one of the simulations for this work was performed using computing resources provided by the Flatiron Institute. 
B.B.~is grateful for funded support by the Simons Foundation, Sloan Foundation, and the Packard Foundation.
Support for V.S.~was provided by NASA through the NASA Hubble Fellowship grant HST-HF2-51445.001-A awarded by the Space Telescope Science Institute, which is operated by the Association of Universities for Research in Astronomy, Inc., for NASA, under contract NAS5-26555.
C.F.~acknowledges funding provided by the Australian Research Council (Future Fellowship FT180100495), and the Australia-Germany Joint Research Cooperation Scheme (UA-DAAD). We further acknowledge high-performance computing resources provided by the Australian National Computational Infrastructure (grant~ek9) in the framework of the National Computational Merit Allocation Scheme and the ANU Merit Allocation Scheme, and by the Leibniz Rechenzentrum and the Gauss Centre for Supercomputing (grant~pr32lo). A.L.R acknowledges support from NASA through
Einstein Postdoctoral Fellowship grant number PF7-
180166 awarded by the Chandra X-ray Center, which
is operated by the Smithsonian Astrophysical Observatory for NASA under contract NAS8-03060; and support
from Harvard University through the ITC Post-doctoral Fellowship.
The software used in this work was in part developed by the DOE-supported Flash Center for Computational Science at the University of Chicago.
The authors acknowledge the use of the following software: yt \citep{Turk11a}, flash \citep{Fryxell2000}, SciPy \citep{SciPy}, scikit-learn \citep{scikit-learn}, Matplotlib \citep{Hunter2007}, astropy \citep{Astropy-Collaboration13a}.

\bibliography{bibliography.bib,vs.bib}{}

\end{document}